\documentstyle[aps,multicol,
epsfig]{revtex}

\begin{document}
\title{\bf Turbulence in a free surface}
\author{ W. I. Goldburg$^1$, J. R. Cressman$^1$, Z. V\"or\"os$^1$, B.
Eckhardt$^2$, and J. Schumacher$^2$}
\address{$^1$ Department of Physics and Astronomy, University of
Pittsburgh, Pittsburgh, PA 15260\\
$^2$ Fachbereich Physik, Philipps-Universit\"at Marburg,
D-35032 Marburg, Germany} \date{\today} \maketitle

\begin{abstract}
We report an experimental and numerical study of turbulent fluid
motion in a free surface.  The flow is realized experimentally on the
surface of a tank filled with water stirred by a vertically
oscillating grid positioned well below the surface.  The effect of
surface waves appears to be negligible so that the flow can
numerically be realized with a flat surface and stress-free boundary
conditions.  The surface flow is unconventional in that it is not
incompressible and neither energy nor enstrophy are conserved.
Nevertheless, according to both experiment and numerical simulation,
the second order structure function $S_2(R)$ scales essentially as for
a three-dimensional system.  However, the surface flow seems to be
more intermittent.
\end{abstract}

\begin{multicols}{2}
Statistically stationary turbulence is readily produced in a tank of
water by vertically oscillating a grid below the water surface.  The
properties of this three-dimensional volume turbulence have been
studied extensively \cite{matsunaga99}.  Particles floating on the
free surface are driven by the horizontal component of the turbulent
flow below.  While this is a classic problem, generally studied in an
oceanographic context where the ``particles'' are micro organisms,
such as phytoplankton \cite{gower81}, or Lagrangian drifters
\cite{stommel49,Davis91}, controlled laboratory measurements of the
relative velocity of floating particles do not appear to exist,
despite a very large number of oscillating grid experiments which
probe the fluid flow in the bulk and even near the surface
\cite{matsunaga99}.  Such surface flows also seem to have received
little attention by those interested in the fundamental aspects of two
or three dimensional turbulence \cite{frisch,lesieur}.  Our
experimental and theoretical results demonstrate that these
two-dimensional surface flows have unusual properties that make them
interesting in connection with scalar transport.  The work opens the
possibility of studying certain aspects of passive scalar dynamics in
divergent flows beyond the ones used for the Kraichnan model which are
Gaussian and white in time \cite{gawedzki00}.  Their properties are
also needed for a hydrodynamic modelling of the distribution of
flotsam driven by underlying turbulence or chaos \cite{maryland}.  Our
study is also related to recent work on quasi two-dimensional
turbulence, which is probed by tracking the motion of floating
particles \cite{jullien99}. The surface flows being comprised of
particles constrained to the surface in both the experimental and
numerical work, are qualitatively different from the motion
of the driving fluid.

Although the fluid is incompressible,
the inability of the floaters to enter the bulk flow assures that they 
will have a
non-zero two-dimensional divergence. The compressibility of the
floating particles causes them to clump, an affect seen in the
numerical and experimental realizations. The main differences between
their motion and that of the underlying fluid comes from the
possibility that the floaters can exchange kinetic energy and
enstrophy with the water molecules below. Hence, the energy and enstrophy 
of the
floaters still would
not be constants of the motion, even if the underlying
fluid were invisid and undriven \cite{kraichnan}.  With the above
conservation laws absent, one lacks the traditional dimensional
arguments to estimate the scaling forms of the velocity and vorticity
structure functions, $S_{2}(R)$ and $S_{\omega}(R)$.  The structure
functions are the main focus of our analysis and are defined as
%-------------
\begin{equation}
S_{2}(R)=\langle\lbrack({\bf v}({\bf x}+{\bf R})-{\bf v}({\bf
x}))\cdot {\bf R}/R\rbrack^{2}\rangle
\end{equation}
%-------------
and 
%-------------
\begin{equation}
S_{\omega}(R)=\langle \lbrack\omega ({\bf x}+{\bf R})-\omega ({\bf
x})\rbrack^{2} \rangle \,,
\end{equation}
%--------
where ${\bf R}$ is the separation between two points and both ${\bf
x}=\{x,y\}$ and ${\bf R}$ are in a horizontal plane.  The vorticity,
of magnitude $\omega (x,y)$, is perpendicular to that plane.  A third
quantity characterizing the surface flow is the dimensionless
compressiblity coefficient ${\cal C}$, defined as
 %---------
\begin{equation}
{\cal C}=\langle({\bf {\nabla}} \cdot {\bf v})^{2}\rangle
/\langle({\bf \nabla}{\bf v})^{2}\rangle\,,
\label{cfactor}
\end{equation}
%----------
which lies between 0 and 1 if the turbulence is isotropic
\cite{gawedzki00}.  As usual, $\langle\cdots\rangle$ denotes an
ensemble average.

The measurements were made in a square plexiglass tank filled with
water to a height that was varied.  The bars of the grid were PVC and
were square in cross section with dimensions of 1 cm.  The grid is
also square and conforms closely with the square shape of the
plexiglass box.  To minimize flexing of the grid when it is
oscillating, a number of the PVC bars are replaced by brass bars.  For
some of the measurements, a vertical post was placed at the center of
the square grid to further suppress flexing, thereby reducing the
amplitude of the surface waves.  The apertures of the grid are also
square, their dimensions being $L$=3.66 cm.  The grid is supported
by four vertical rods which run through pillow blocks, assuring its
smooth vertical motion.  All the parameters used in our set-up were
typical of prior oscillating grid experiments \cite{matsunaga99},
except that the lateral dimensions of the tank were somewhat
larger.The grid is driven sinusoidally from above by a 1/4 hp motor.
The vertical amplitude of grid motion was $A=$ 1.25 cm.  Whereas
these parameters were varied by only a small amount, the distance $Z$
between the grid and the water surface (at $z_0$=0) was rather
widely varied.  Reducing $Z$ increases the Reynolds number of the
turbulence on the surface and in the bulk close to the surface
($z_0>0$).  For all the measurements presented here, the oscillation
frequency $f$ of the grid was 4.5 Hz $\pm$ 0.3 Hz.

The flow was measured using the technique of particle imaging
velocimetry (PIV).  There one illuminates the seed particles with a
(horizontal) sheet of laser light and tracks their motion
photographically.  A CCD camera was centered at a distance that was 23
cm from the nearest wall.  It was mounted above the tank where we
could capture images of the surface particles or those in the
interior.  The particle illumination was furnished by a Nd-Yag laser,
with the beam spread into a horizontal sheet by a cylindrical lens.
This sheet of light was typically 10 cm wide and 0.5 cm thick.  The
floaters were particles of various diameters ranging from 10 to 50
$\mu$m.  The interior measurements were made using neutrally buoyant
polystyrene spheres of diameter 10 $\mu$m.  The images were
acquired using a commercial particle imaging velocimetry apparatus
\cite{TSI} and the particle tracking was done using in-house software.
To form a statistically reliable ensemble average, each measurement of
$S_{2}(R)$ lasted several minutes and included $\sim$ 200 image pairs.
Each $S_{2}(R)$ was determined from roughly a million particle pairs.

%-------------------------------------------------------------
\begin{figure}
\begin{center}
\epsfig{file=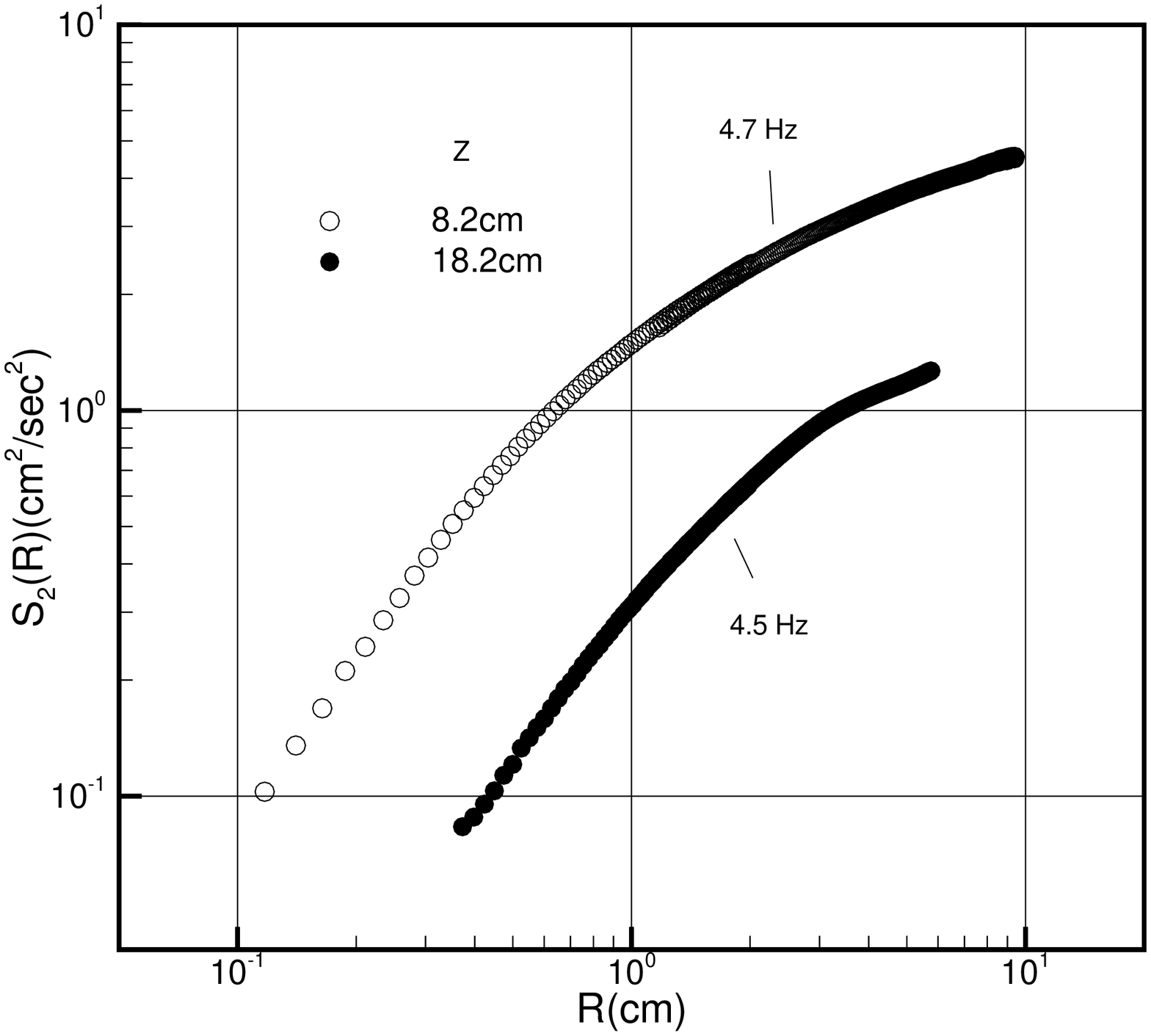,width=8.5cm,angle=-90}
\end{center}
\end{figure}

\vspace{-3cm}
FIG. 1. Log-log plot of the second moment of the (longitudinal)
velocity difference at the indicated distances ($Z$) between the
surface and the oscillating grid.  The oscillating frequency of the
grid and the surface-to-grid distances, $Z$, are indicated.  Other
parameters are $Z=8.2$ cm: $Re_{\lambda}$=80, $l_{0}$=3.6 cm,
$v_{rms}$=1.27 cm/s; $Z$= 18.2 cm: $Re_{\lambda}$=50, $l_{0}$=1.5 cm,
$v_{rms}$=0.7 cm/s.  For both sets of measurements
${\cal C}$=0.5 $\pm$ 0.1.

\vspace{0.5cm}
\noindent
%-------------------------------------------------------------
The surface and bulk turbulence are characterized by the outer scale
of the turbulence $l_{0}$ and the Taylor microscale $\lambda$.  The
outer scale is defined as the integral of the normalized velocity
autocorrelation function $C(R)=\langle v(x+R) \cdot v(x)\rangle
/v_{rms}^{2}$, where the brackets indicate an average over points
$x,y$ and over many images.  Here $v_{rms}^{2} \equiv \langle
(v_{i}(x,y)-\langle v_{i}(x,y)\rangle)^{2}\rangle$ where $i$ is one of
the two arbitrarily chosen directions in the horizontal plane.
Separate experiments established that the turbulence is almost
isotropic in the horizontal plane (see below).  The Taylor microscale
is defined here as $\lambda =
\sqrt{(v_{rms}^{2})/\langle(dv(x)/dx)^{2}\rangle}$.  The Taylor
microscale Reynolds number is $Re_{\lambda}=v_{rms}\lambda/\nu$,
where $\nu$ is the kinematic viscosity (in water $\nu$=0.01
cm$^{2}$/s).  The surface motion is further characterized by the
vertical component of the vorticity $\omega$ and by the dimensionless
compressibility coefficient ${\cal C}$ [see Eq.~(\ref{cfactor})].

Figure 1 is a log-log plot of $S_{2}(R)$ in the surface measured for
$Z$= 18.2 cm (closed circles) and for $Z$= 8.2 cm (open circles).
The grid oscillation frequency for the two sets of measurements are
indicated.  The outer scale of the turbulence $l_{0}$ at $Z$=18.2
cm and 8.2 cm are 1.5 and 3.6 cm, respectively.  The Taylor microscale
Reynolds numbers $Re_{\lambda}$ are 80 ($Z$ = 8.2 cm) and 50 ($Z$= 18.2 cm).
Technical limitations blocked measurement of $S_{2}$ at $R$
less than 1 mm.  The measurements in Fig.~1 suggest that $S_{2}(R)$ is
approaching its saturation value $2v_{rms}^{2}$ at large $R$.  The
more interesting case is $S_{2}$ at smaller $R$, where one finds that
$S_{2}(R) \propto R^{\zeta}$, with $\zeta$=1.6 and 1.4 at $Z$= 8.2
and 18.2 cm, respectively.  The dissipative range at even smaller
$R$, where $S_{2}(R) \rightarrow R^{2}$, was not accessible in the
laboratory experiment \cite{frisch}.  For these, and the other surface
measurements, the dimensionless compressibility ${\cal C}$ defined
above, was 0.5 $\pm$ 0.1.  Reynolds numbers in the experiment and the
simulation are too small to develop a significant interval with
algebraic scaling in $S_2 (R)$ between the dissipative, $S_{2}(R)
\propto R^{2}$, and the saturation range $R\gg l_{0}$, where $S_{2}(R)
\propto R^0$.  Therefore the extended self similarity (ESS)
representation is used \cite{benzi93} and a local scaling exponent
$D_{2,\,3}(R)= d \log_e S_{2}(R)/d \log_e G_{3}(R)$ for all $R$ in
the interval 0.1 cm $ < R <$ 7 cm is found to give $0.70 \pm 0.10$
(see upper panel of Fig.~2).  For the ESS plots, the third order
structure function $ G_3(R)= \langle\left|({\bf v}({\bf x}+{\bf
R})-{\bf v}({\bf x}))\cdot {\bf R}/R\right|^{3}\rangle$ is needed.

Since the surface is entirely driven by the grid-generated turbulence
in the water, it is important to measure $S_{2}(R)$ below the surface.
We have made such measurements at distances $z_0$ in the range 0.5 to
2 cm and for various grid-to-surface distances $Z$.  At $Z$  9.6 cm,
$z_0$= 2.0 cm, the second moment is very well fitted to a power law
form in the interval 0.1 cm $ < R <$ 4 cm.  In this interval $\zeta$=0.9 $\pm$ 0.2,
which is somewhat larger than the  Kolmogorov
value of 2/3 \cite{frisch}.  Other relevant parameters for this bulk
measurement are $Re_{\lambda}$= 100 and $l_{0}$= 3.1 cm.  The
corresponding parameters measured on the surface are $Re_{\lambda}$= 40
and $l_{0}$= 3.5 cm.  At all values of $R$, the bulk value of the
second moment is roughly an order of magnitude larger than $S_{2}(R)$
measured at the surface.  A dimensional argument suggests the
crossover between bulk and surface behavior should take place at a
depth $\delta z$ below the surface given by $\delta z \simeq
\nu/{\tilde v}$ where ${\tilde v}$ is the bulk rms value of $v_{z}$.
This gives $\delta z \simeq$ 0.1 mm.  The numerical studies \cite{ES00} indicate that the
prefactor in this dimensional estimate can be as large as 10, making
$\delta z \simeq$ 1 mm, which is still too close to the surface for us
to measure.

%-------------------------------------------------------------
\begin{figure}
\begin{center}
\epsfig{file=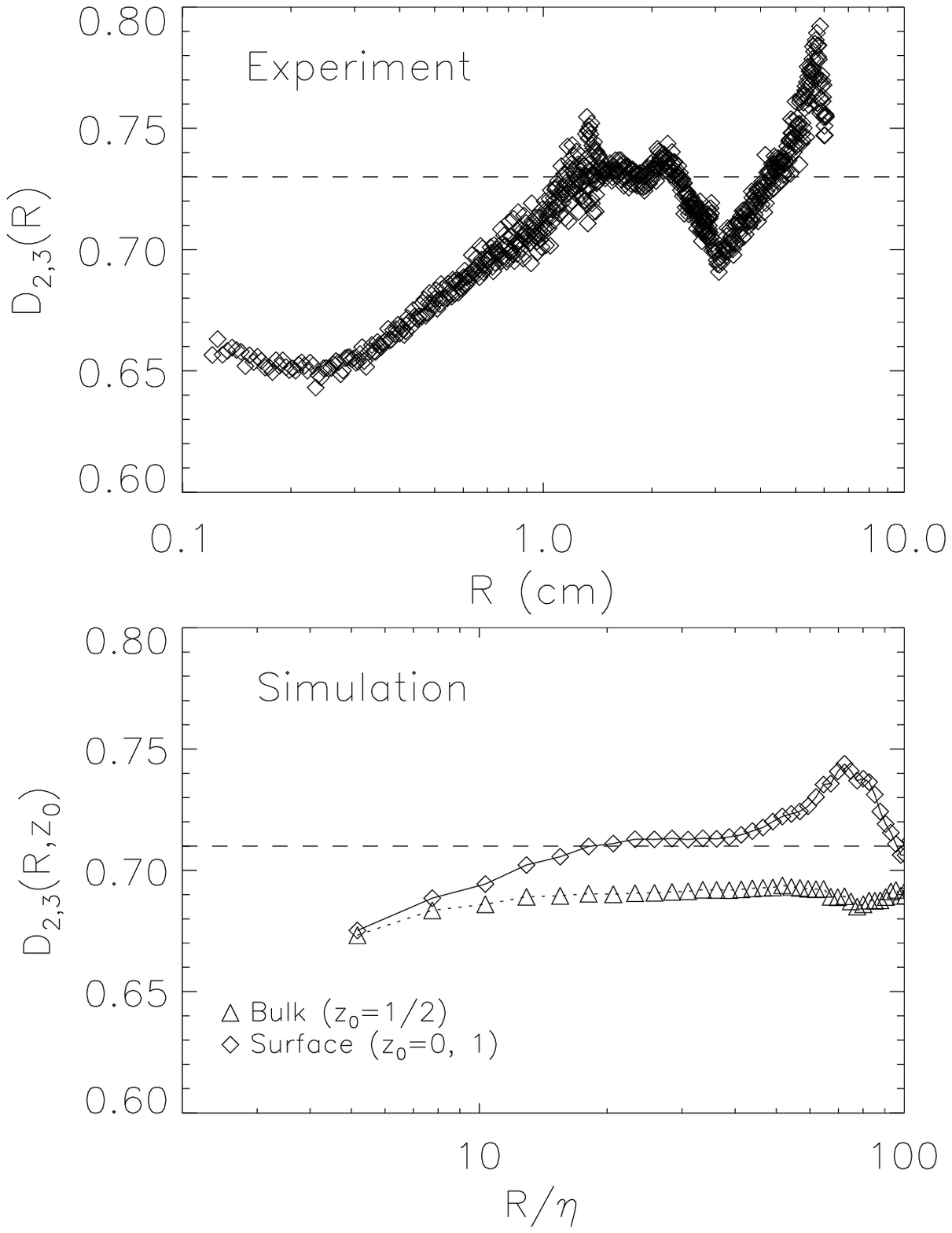,width=7cm}
\end{center}
\end{figure}
FIG. 2. Local scaling exponent, $D_{2,3}(R)$, of an extended self
similarity plot of second order moment vs.  absolute third order
moment.  Upper panel: The parameters for this experiment are the same
as in Fig.~1 at $Z$=18.2 cm.  Lower panel: Comparison of
$D_{2,3}(R)$ in the bulk at $z_0$=0.5 (triangles) with that on the
two surfaces at $z_0$=0 and 1 (diamonds).  Data were taken from a
numerical simulation at $Re_{\lambda} \simeq$ 100.  Dashed lines in
both panels are drawn at $D_{2,3}(R)$=0.73 and 0.71, respectively.

\vspace{0.5cm}
\noindent
%-------------------------------------------------------------
We consider the possibility that the surface velocity fluctuations
originate from gravity and capillary waves generated by the 4.4 Hz
oscillation frequency of the grid.  Even though the amplitude of the
surface waves was rather small ($A \simeq $ 0.5 mm) it is comparable
to the thickness of the boundary layer near the surface.  The
appropriate wave turbulence theory treats the nonlinear coupling of
surface modes $\Omega(k)$ of different wavenumbers $k$
\cite{zakharovetal}.  This theory is of Hamiltonian form, which
requires that the (2D) velocity field in the theory is obtained from a
velocity potential, so that the vorticity of the surface particles is
zero.  Measurements of the (vertically-directed) vorticity of the
surface particles show that it is quite large.  Figure~4 is a log-log
plot of the second moment of the surface vorticity fluctuations,
$S_{\omega}(R)$, for a grid-to-surface distance $Z$= 8.5 cm.  This
function saturates at a rather small value of $R$ and approaches the
limiting value $2\langle\omega^{2}\rangle$.  A dimensionless measure
($Q$) of the surface vorticity is ${S_{\omega}(R)} R^2/{S_{2}(R)}$.
At $R$ equal to the outer scale of the turbulence, $l_{0} \simeq 3$
cm, $Q \simeq$ 1, which suggests that it is probably too large to
satisfy the zero-vorticity requirement of wave turbulence theory, even
though this theory successfully accounts for the motion of particles
floating on a container of vertically oscillating fluid where
capillary waves are excited \cite{schroder96,schroder98}.

The experimental observations are, however, in good agreement with a
model for turbulence in a flat surface bounding a three-dimensional
volume with fully developed turbulence governed by the incompressible
Navier-Stokes equation \cite{ES00}.  This fluid is in a computational
box with periodic boundary conditions applied to the lateral edges but
with the fluid particles on the upper and lower surfaces obeying
free--slip boundary conditions, with $v_{z}=\partial_z
v_x=\partial_z v_y=0$.  The equations were solved by means of a
pseudospectral method with resolution $N_x\times N_y\times N_z$ of
$256\times 256\times 65$.  A statistically steady state of the
turbulent flow was achieved by forcing in a wavenumber range
comparable to the inverse of the box height (for more details see
\cite{SE00}).  Here $Re_{\lambda} \simeq$ 100.

%-------------------------------------------------------------
\begin{figure}
\begin{center}
\epsfig{file=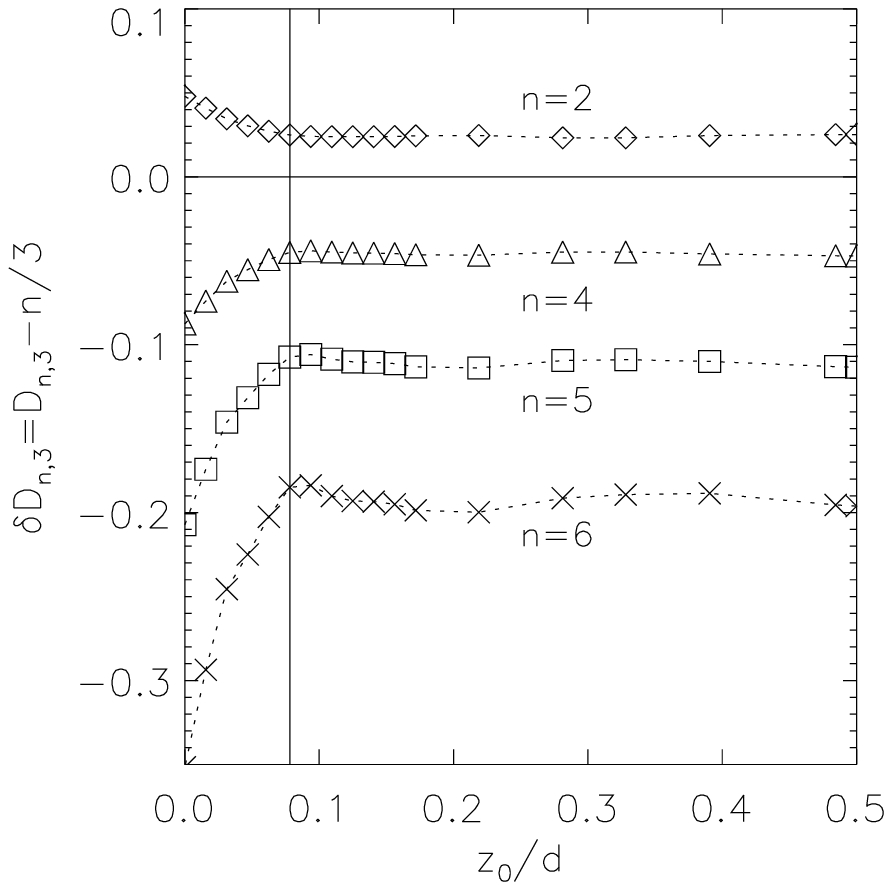,width=7cm}
\end{center}
\end{figure}
FIG. 3. Deviations of the local extended self-similarity (ESS) slope
from the classical Kolmogorov scaling for different heights $z_0$ of
the averaging plane and for different orders $n=2$ to $n=6$.  All
data points are calculated from $2\times 114$ statistically
independent samples at $Re_{\lambda} \simeq$ 100 and by averaging
$D_{n,\,3}(R,z_0)$ for scales $R$ between $18\eta$ and $41\eta$.

\vspace{0.5cm}
\noindent
%-------------------------------------------------------------
The local ESS scaling exponent $D_{2,\,3}(R)$ lies in the range
between 0.67 and 0.74 in the interval 5 $< R/\eta <$ 100 with the
Kolmogorov length $\eta=(\nu^3/\epsilon)^{1/4}$ (lower panel of
Fig.~2) .  This result is in good agreement with the experimental data
also shown in Fig.~2 (upper panel).  The simulation results show a
plateau between $18\eta$ and $41\eta$ for the local scaling exponents
in both cases, at the surface and in the bulk.  Intermittency
corrections in the surface seem to be stronger than in the bulk.  In
Fig.~3, a systematic investigation of these deviations from the
classical Kolmogorov scaling with respect to $z_0$, the vertical
position of the horizontal plane of averaging is presented.  Our
results indicate increasing deviations toward the free surface.  The
two-dimensional surface compressibility ${\cal C}$ in the calculation
is 0.5, in excellent agreement to the laboratory-measured value of
$0.5\pm 0.1$. This value of ${\cal C}$ is expected for isotropic
turbulence.

%-------------------------------------------------------------
\begin{figure}
\begin{center}
\epsfig{file=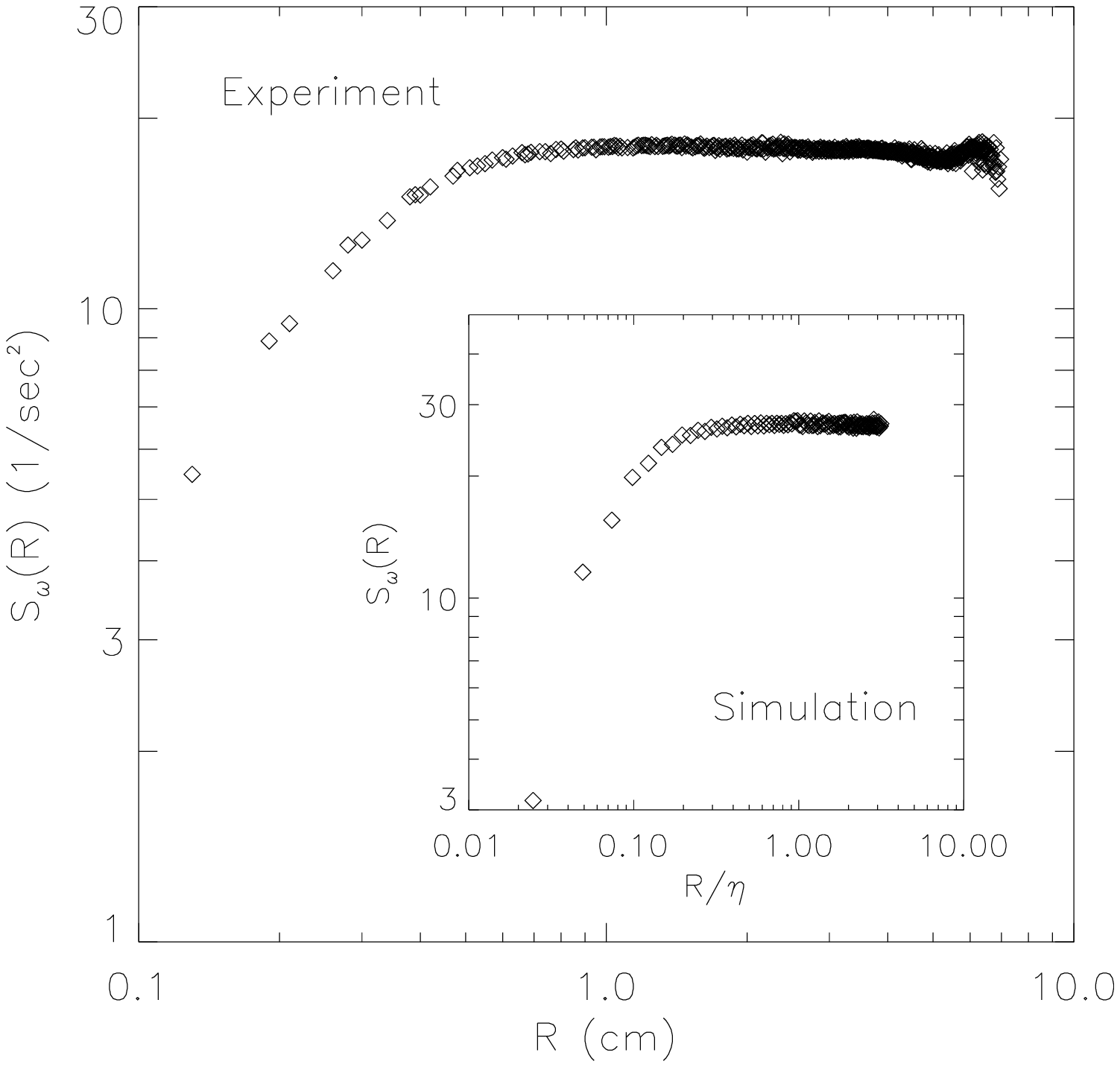,width=7cm}
\end{center}
\end{figure}
FIG. 4. Second moment of vorticity differences for the floating
particles, at $Z$=9.6 cm.  Here $f$=4.7 Hz, $Re_{\lambda}$=35,
$l_{0}$=1.7 cm, and $v_{rms}$=0.67 cm/s.  Again ${\cal C} \simeq$
0.5.  Inset: Model calculation of $S_{\omega}(R)$ at the free-slip
surfaces, with $Re_{\lambda}\simeq $ 100.  Here ${\cal C}$=0.5.

\vspace{0.5cm}
\noindent
%-------------------------------------------------------------
The above model was also applied to the calculation of
$S_{\omega}(R)$, and the results are presented in the inset of Fig.~4.
Both the calculated and measured $S_{\omega}(R)$ are qualitatively
similar and show that this function rises rapidly with increasing $R$
in these log-log plots and reaches its saturation value,
2$\langle\omega^{2}\rangle$ at a rather small value of $R$\,.

In summary, we have described a laboratory experiment and a model
calculation for particles that float on a sea of three-dimensional
turbulence.  Even though the surface particles move in a plane, there
appears to be no basis for describing the floating particle system in
terms of the laws of two-dimensional turbulence or of turbulent
surface waves.  Important characteristics of the floating particle
system are its large compressibility, large vorticity and the
absence of conservation laws for energy and enstrophy.  The
measurements and calculations presented here indicate the existence of
an inertial range where $S_2(R)$ scales as $R^{\zeta}$, with $\zeta $
quite close to 2/3, as in 3D flows and in the inverse cascade of 2D
turbulence.

In this work we have profited from interactions with P. Alstr\o m, S.
Banerjee, D. Lohse, C. Cheung, G. Falkovich, V. Horv\'{a}th, M.
Rivera, E. Schr\"oder, X. L. Wu, and M. Vergassola.  The particle
tracking software of M. Rivera was very important to the success of
this experiment.  Derek Hohman provided valuable assistance with the
experiment.  The work has been strongly influenced by the attendance
of two of us (W.I.G., B.E.) at the Institute for Theoretical Physics
at Santa Barbara.  Research support from NASA, the National Science
Foundation, and the European Community is gratefully acknowledged.  We
thank the John von Neumann--Institut f\"ur Computing at the
Forschungszentrum J\"ulich for support and computing time on a Cray
T-90 for the numerical simulations.

\end{multicols}
\end{document}